\documentclass[preprint,superscriptaddress]{revtex4-1}
\usepackage{bibentry,natbib}
\usepackage{latexsym,bm}
\usepackage{graphicx}
\usepackage{rotating}
\usepackage{amsmath}
\usepackage{color}
\DeclareGraphicsExtensions{.jpg}

\begin{document}
\title{Theoretical study on the hyperfine interaction constants and the isotope shift factors for the 3s$\bm{^{2}$ }  \bm {$^{1} S_{ 0}$ $-$ $ {3s3p}$}  \bm{$^{3,1}P^o_1$} transitions in Al$^{\bm +}$ }

\author{Tingxian Zhang}
\affiliation{College of Physics and Electronic Engineering, Northwest Normal University, Lanzhou 730070 P. R. China}
\affiliation{Institute of Applied Physics and Computational Mathematic, Beijing 100088, P. R. China}
\author{Luyou Xie}
\affiliation{College of Physics and Electronic Engineering, Northwest Normal University, Lanzhou 730070 P. R. China}
\author{Jiguang Li}
\email{li$_$jiguang@iapcm.ac.cn}
\affiliation{Institute of Applied Physics and Computational Mathematic, Beijing 100088, P. R. China}
\author{Zehuang Lu}
\affiliation{MOE Key Laboratory of Fundamental Physical Quantities Measurement, School of Physics, Huazhong University of Science and Technology, Wuhan 430074, P. R. China}

\begin{abstract}
 We calculated the magnetic dipole and the electric quadrupole hyperfine interaction constants of 3s3p $^{3,1}P^o_1$ states and the isotope shift, including mass and field shift, factors for transitions from these two states to the ground state 3s$^2~^1S_0$ in Al$^+$ ions using the multiconfiguration Dirac-Hartree-Fock method. The effects of the electron correlations and the Breit interaction on these physical quantities were investigated in detail based on the active space approach. It is found that the CC and the higher-order correlations are considerable for evaluating the uncertainties of the atomic parameters concerned. The uncertainties of the hyperfine interaction constants in this work are less than 1.5\%. Although the isotope shift factors are highly sensitive to the electron correlations, reasonable uncertainties were obtained by exploring the effects of the electron correlations. Moreover, we found that the relativistic nuclear recoil corrections to the mass shift factors are very small and insensitive to the electron correlations for Al$^{+}$. These atomic parameters present in this work are valuable for extracting the nuclear electric quadrupole moments and the mean-square charge radii of Al isotopes.

\end{abstract}
\maketitle

\section{Introduction}
$~$ The exotic nuclei and the nuclei close to the dripline have some peculiar structures and properties, such as larger nuclear radii and reaction cross section, etc. Investigating the properties of these nuclei, especially for the evolutionary trend of the nuclear properties along the isotope chain, is helpful for understanding the many-body interactions within the nucleus and improving the nuclear structure theory~\cite{Campbell2016}. To characterize the nuclear properties quantitatively, nuclear electric quadrupole moment $Q$ and mean-square charge radius $\langle$$r^2$$\rangle$ are required. In the chain of aluminum isotopes, the proton-rich isotope $^{23}$Al has the proton-halo structure~\cite{Zhang2002,Nagatomo2010}, the neutron-rich isotopes $^{31-33}$Al are in the vicinity of the ``island of inversion", and the $^{26}$Al is a self-conjugate nucleus~\cite{Cooper1996}. Up to date, the $Q$ and  $\langle$$r^2$$\rangle$ values are only available for a few isotopes of aluminum, and the accuracies are not high enough except for $^{27}$Al~\cite{Kello1999,Sundholm1992,Angeli2011}. For example, the $Q$ value of $^{26}$Al, with an uncertainty about 12\%, was deduced from the experimental hyperfine structures measured by using the atomic laser spectroscopy~\cite{JMG1997} in assistance of the relation $^{26}Q/^{27}Q =~^{26}B/^{27}B$ ($B$ is the electric quadrupole hyperfine interaction constants). For the $^{23}$Al, $^{28}$Al, and $^{31-33}$Al isotopes, the nuclear electric quadrupole moments were measured by the $\beta$-ray-detected nuclear magnetic resonance ($\beta$-NMR) or $\beta$-ray-detected nuclear quadrupole resonance ($\beta$-NQR) method~\cite{Nagatomo2010,Stockmann1978,Nagae2009,Kameda2007,Gaudefroy2009,DeRydt2009}. However, the error bars, about 20\% for $^{23}$Al and $^{31}$Al, are so large that the nuclear deformation parameter $\beta$ cannot be determined. Compared with the nuclear quadrupole moment, the nuclear charge mean-square-root is scarce. The $\langle$$r^2$$\rangle$ value of these interesting nuclei has not been reported. As a consequence, it is indispensable to determine accurately the $Q$ and $\langle$$r^2$$\rangle$ values of aluminium isotopes for exploring their nuclear properties.

$~$ The nuclear-model-independent data $Q$ and $\langle$$r^2$$\rangle$ can be extracted by combining the measured hyperfine structures and isotope shifts
with theoretical prediction of the electric field gradient and the isotope shift factors. The collinear laser spectroscopy technique has been developed to measure the isotope shifts and the hyperfine structures of the exotic or radioactive nuclei with short lifetime and low production~\cite{Campbell2016,Be2009}. In addition, the high-precision measurements of hyperfine structures and isotope shifts of Al$^+$ ions can be obtained based on the Al$^+$ ion optical clock~\cite{Lu2016}, and thus can be used to verify the calculation. For these reasons, the Al$^+$ ion can be considered as a good candidate for extracting the nuclear electric quadrupole moment $Q$ and mean-square charge radius $\langle$$r^2$$\rangle$ of Al isotopes.

$~$ Most of the earlier theoretical works related to the 3s3p $\rm{^{1,3}P_1^o}$ $-$ 3s$^{\rm 2}$ $\rm{^1S_0}$ transitions in Al$^+$ ions focused on transition energies and probabilities~\cite{Konovalova2015,Santana2015,Stanek1996,Safronova2000,Safronova20002,Laughlin1979,hibbert1987,Brage1998,Jonsson1997,Zou2000,Zou2001,FroeseFischer2006,Kang2009,Kang2010,Andersson2010}. For the hyperfine interaction constants we only found theoretical reports by Kang \textit{et al.} and Andersson \textit{et al.}. In their works the off-diagonal magnetic dipole and electric quadrupole hyperfine interaction constants of the 3s3p $^{1,3}P_1^o$ states were calculated for investigating the effect of hyperfine interaction on the lifetime of metastable states 3s3p $^3P^o_{0,2}$~\cite{Kang2009,Andersson2010}. While the isotope shift factors of the transitions 3s3p $^{1,3}P^o_1$ $-$ 3s$^2~^1S_0$ for Al$^{+}$ ions have not been reported.

$~$ In this work, we calculated the hyperfine interaction constants of 3s3p $^{3,1}P^o_1$ states and the mass shift and field shift factors of the 3s$^2$  $^1S_0$ $-$ 3s3p  $^{3,1}P^o_1$ in Al$^+$ using the multi-configuration Dirac-Hartree-Fock (MCDHF) method. The active space approach was adopted to investigate the effects of electron correlations on the hyperfine interaction constants and the isotope shift factors in detail. Based on this, we built a computation model, which can capture electron correlations effectively and allow us to obtain high-precision atomic parameters concerned. In addition, the relativistic nuclear recoil corrections to the isotope shift factors were discussed. We hope this work could support the experimental investigations about the Al$^+$ ion optical clock and the studies about the nuclear properties of Al isotopes.

\section{Theoretical method}
\subsection{MCDHF method}
In the MCDHF method an atomic state wavefunction (ASF) $\Psi $ is a linear combination of configuration state functions (CSFs) $\Phi$ with the same parity $P$, total angular momentum $J$ and it's component along the $z$ direction $M_J$\cite{Grant2007},
\begin{equation}
\label{MCDHF}
 \Psi (\gamma PJM_J) = \sum\limits_{i=1}^{N_{\rm CSF}} {{c_i}\Phi ({\gamma _i}PJM_J)}.
\end{equation}
In the Eq~\eqref{MCDHF} $c_i$ is the mixing coefficient, and $\gamma$ represents other appropriate labeling of the CSF. The CSFs are the linear combinations of products of one-electron Dirac orbitals. In the self-consistent field (SCF) calculation both the mixing coefficients and the orbital are optimized. After the virtual orbital set is obtained more electron correlations can be included in the relativistic configuration interaction (RCI) calculations by further expanding the configuration space. In the RCI calculations all orbitals are kept frozen, and only the mixing coefficients are variable. The Breit interaction in the low-frequency approximation,

\begin{equation}
B_{ij}=-\frac{1}{2r_{ij}} \bigg[\bm{\alpha}_i \cdot \bm{\alpha}_j+\frac{(\bm{\alpha}_i\cdot \bm{r}_{ij})(\bm{\alpha}_j\cdot \bm{r}_{ij})}{r_{ij}^2}\bigg]
\end{equation}
is also taken into account in the RCI computation~\cite{LJG2012}.

\subsection{Hyperfine interaction}
The hyperfine interaction is caused by the interaction between the electrons and the electronmagnetic multipole moments of the nucleus, and its Hamiltonian can be represented using the spherical tensor operators $\bm{T}^{(k)}$ and $\bm{M}^{(k)}$~\cite{Lindgren1983},
\begin{equation}
H_{\rm hfs}=\sum_{k\geq1}\bm{T}^{(k)}\cdot\bm{M}^{(k)}.
\end{equation}
Here, $k=1$ and $k=2$ represent the magnetic dipole and electric quadrupole interactions, respectively, and the higher-order terms are tiny and neglected in this work. For a $N$-electron atom the electronic tensor operators $\bm{T}^{(1)}$ and $\bm{T}^{(2)}$ are the sums of the one-electron operators,
\begin{equation}
\bm{T^{(1)}}=\sum_{j=1}^N\bm{t}^{(1)}(j)=\sum_{j=1}^N-i\alpha \Big(\bm{\alpha}_j\cdot{ \bf l }_j\bm{C}^{(1)}(j)\Big)r_j^{-2}
\end{equation}
and
\begin{equation}
\bm{T^{(2)}}=\sum_{j=1}^N\bm{t}^{(2)}(j)=\sum_{j=1}^N-\bm{C}^{(2)}(j)r_j^{-3}.
\end{equation}
Here, $i$ is the imaginary unit, $\alpha$ is the fine-structure constant, $\bm{C}^{(1)}$ and $\bm{C}^{(2)}$ are spherical tensor operators, and $\bf{l}$ is the orbital angular momentum operator. The nuclear tensor operators $\bm{M}^{(1)}$ and $\bm{M}^{(2)}$ are related to magnetic dipole moment $\mu_I$ and electronic quadrupole moment $Q$ of the nucleus with spin $I$ through~\cite{Parpia1996}:
\begin{equation}
\mu_I=\langle IM_I(=I)|\bm{M}_0^{(1)}|IM_I(=I)\rangle
\end{equation}
and
\begin{equation}
Q=\langle IM_I(=I)|\bm{M}_0^{(2)}|IM_I(=I)\rangle.
\end{equation}
In the first-order perturbation approximation, the magnetic dipole and the electric quadrupole hyperfine interaction constants $A_J$ and $B_J$ are~\cite{Parpia1996}
\begin{equation}
A_J=\frac{\mu_I}{I}\frac {1}{[J(J+1)]^{1/2}}\langle\Psi(PJ)\|\bm{T}^{(1)}\|\Psi(PJ)\rangle
\end{equation}
and
\begin{equation}
B_J=2Q\bigg[\frac{J(2J-1)}{(J+1)(2J+3)}\bigg]^{1/2}\langle\Psi(PJ)\|\bm{T}^{(2)}\|\Psi(PJ)\rangle.
\end{equation}
\subsection{Isotope shifts}
The isotope shift is composed of the mass shift (MS) and the field shift (FS). The former is caused by the motion of nucleus with finite mass and the latter by the nuclear distribution. For a transition the isotope shift is
\begin{eqnarray}
\Delta E_{\rm IS}^{A,A^\prime}&= &\Delta E_{\rm MS}^{A,A^\prime}+\Delta E_{\rm FS}^{A,A^\prime}\\
\label{123}
       &=&\bigg(\frac1{M}-\frac1{M^\prime}\bigg)\Delta K_{\rm RMS}+\Delta F \delta\langle r^2\rangle^{A,A^\prime}.
\end{eqnarray}
In the formulas above, the $M$ and $M^\prime$ are the nuclear masses of the isotopes $A $ and $A^\prime$, respectively. In addition, $\delta\langle r^2\rangle^{A,A^\prime}\equiv \langle r^2\rangle^A -\langle r^2 \rangle^{A^\prime}$. The mass shift (MS) can be separated into the normal mass shift (NMS) and the specific mass shift (SMS). The relativistic nuclear recoil Hamiltonian $H_{\rm RNMS}$ and $H_{\rm RNMS}$ correspond to one-body and two-body relatively mass shift operator, respectively~\cite{Palmer1987,Torbohm1985},
\begin{eqnarray}
H_{\rm RMS}&=& H_{\rm RNMS}+H_{\rm RSMS}\\
      &=&\frac1{2M}\sum _{i}\{\bm{p}_i^2 -\frac{\alpha z}{r_i} [\bm{\alpha}_i+\frac{(\bm{\alpha }_i\cdot \bm{r}_i)\bm{r}_i}{r_i^2}]\cdot \bm{p}_i\}\\
 &&+\frac1{2M}\sum _{i\neq j}\{\bm{p}_i\cdot \bm{p}_j-\frac{\alpha z}{r_i} [\bm{\alpha}_i+\frac{(\bm{\alpha}_i\cdot \bm{r}_i)\bm{r}_i}{r_i^2}]\cdot \bm{p}_j\}.
\end{eqnarray}
For a given level, the normal mass shift factors $K_{\rm RNMS}$ and the specific mass shift factors $K_{\rm RSMS}$ are defined as~\cite{LJG2016}
\begin{equation}
K_{\rm RNMS}\equiv M\langle\Psi|H_{\rm RNMS}|\Psi\rangle\\
\end{equation}
and
\begin{equation}
K_{\rm RSMS}\equiv M\langle\Psi|H_{\rm RSMS}|\Psi\rangle.
\end{equation}

For a transition, the $\Delta K$ is the difference of the mass shift factors between the upper ($u$) and lower ($l$) levels, and $\Delta K_{\rm RMS}=\Delta K_{\rm RNMS}+\Delta K_{\rm RSMS}$.

The field shift (FS) factor for a given transition is expressed as~\cite{Blundell1987}
\begin{equation}
\Delta F=\frac{2\pi}{3}(\frac{Ze^2}{4\pi\epsilon_0})\Delta|\Psi(\bm{0})|^2,
\end{equation}
where, $\Delta|\Psi(\bm{0})|^2$ is the change of the total electronic probability density at the origin
\begin{equation}
\Delta|\Psi(\bm{0})|^2=\Delta\rho^e(\bm{0})=\rho_u^e(\bm{0})-\rho_l^e(\bm{0}).
\end{equation}
\section{Computational model}
$~$ Generally the precision of the calculated atomic parameters mainly depend on description of electron correlations in the atomic system. In the frame of the MCDHF method one can consider electron correlations systematically adopting the active space approach. According to the perturbation theory the electron correlations can be divided into the first-order and the higher-order correlations. The first-order correlation effects are captured by the CSFs generated through the single (S) and double (D) excitations from the occupied orbitals in the single reference configuration (SR) sets. In order to capture the electron correlations efficiently the occupied orbitals in the reference configuration are separated into the valence and the core orbitals. Therefore, the first-order correlation is composed of the correlation between the valence electrons (VV correlation), the correlation between valence and core electrons (CV correlation), and the correlation between core electrons (CC correlation). The one beyond the first-order correlation is defined as the higher-order correlation and captured by the configuration space expanding from the multireference (MR) configurations set, {\it {i.e.}} the ``MR-SD" model. It was shown that this method is capable of accounting for the electron correlation for the complex ions and atoms~\cite{LJG2012,Livio-Mg,Livio-Al,Carette2013,Carette2010,Andersson2016}.

\subsection{Capture of the first-order correlations}
$~$ In our calculations, the atomic state wave function of even 3s$^2$ and odd 3s3p states were optimized separately. We treated the 1s, 2s and 2p orbitals as the core, and the others as the valence orbitals. The VV and CV correlations were taken into account in the SCF calculations. As shown in the second column of Table \ref{3s2AO} and Table \ref{3s3pAO} the active orbitals (AO) were enlarged layer by layer in order to monitor the convergence of the physical concerned. These steps were labeled by CV$_{1s}$-$nl$, where the subscript $1s$ means the CV correlation between the 1s core and the valence orbitals was involved. $n$ and $l$ is the maximum principal quantum number and the maximum angular quantum number of the outermost active orbitals at each step, respectively. In our test calculations it was found that the contributions from orbitals with large orbital angular momentum $l$ (such as 7i, 7h, 8g orbitals, etc.) are fractional to atomic parameters concerned, so these orbitals were not included in the set of active orbitals. The final set of active orbitals in our calculations was composed of eleven layers of orbitals with $l\leq2$, seven layers with $l=3$, three layers with $l=4$ and one layer with $l=5$. The number of CSFs for 3s$^2$ and 3s3p states in each step are also presented in Table \ref{3s2AO} and Table \ref{3s3pAO}. The SCF calculations started from the Dirac-Hartree-Fock approximation (labeled as DF in the table), in which the occupied orbitals in the reference configurations were optimized as spectroscopic. Subsequently, these orbitals were kept frozen, and only the added orbitals in the active set were variable. At last, the orbitals sets formed in the step CV$_{1s}$-$13d$ were fixed in subsequent RCI calculations in which the CC$_{2s}$ and the higher-order correlations were included.

$~$ The correlation in the $n=2$ core (labeled as CC$_{2s}$) was taken into account by allowing the single and double excitations from the 2s and 2p core orbitals to the largest active set. So far, all the first-order correlation were included in our calculations except the correlation between the core orbitals 1s (CC$_{1s}$ correlation). In fact the CC$_{1s}$ correlation is negligible and will be discussed in section below.\\

\subsection{Capture of the higher-order correlations \label{ho}}
$~$ The higher-order correlation can be captured in two ways. One is to add the CSFs generated by the triple (T) and quadruple (Q) substitutions from the single reference configuration, and the other is to include those produced through the SD substitutions from the multireference configurations. The first way is impracticable for complex atoms, since the configuration space will be expanded too rapidly. Moreover, it is unnecessary to capture all TQ substitutions in practical calculations, because the contributions from most of them are tiny. Actually, the SD excitations from the multireference configurations set is equivalent to the restricted TQ excitations from the single reference configuration, but reserve the important TQ excitations by properly selecting the reference configurations. Therefore, the most important higher-order correlation can be captured by using the MR-SD model. From the physical viewpoint the multi-reference configuration set is composed of the CSFs with large mixing coefficient in the first-order configuration space. In this case, we selected the dominant CSFs in the configuration spaces obtained with the CC$_{2s}$ model to form the multireference configuration set, since the contributions of the CC$_{1s}$ correlations to the mixing coefficient of the dominant CSFs are negligible. The dominant CSFs can be identified according to the weight factor that is defined as
{\setlength\abovedisplayskip{5pt}
\setlength\belowdisplayskip{5pt}
\begin{equation}
\omega =\bigg(\sum_i c_i^2\bigg)^{1/2}.
\end{equation}
Here, the sum extend over the CSFs that belong to the same configuration~\cite{Carette2013}. The Table \ref{MR} shows the weight of the selected configurations. It should be noted that the weight factors in this table were obtained by summing the configuration with $c_i > 0.01$, because the mixing coefficients $c_i < 0.01$ only have tiny effects on the calculated weight factor of the configuration.\\
$~$ In order to explore the convergency of the higher-order correlation effect, we expanded the multireference configurations set in terms of the condition $\omega > 0.05$, $\omega > 0.02$ and $\omega > 0.01$, {\it {i.e.}} the MR$_1$, MR$_2$ and MR$_3$ sets in Table \ref{MR}, respectively. It should be emphasized that only the higher-order correlation between valence orbitals (3s and 3p) and the $n=2$ core orbitals (2s and 2p) was considered in our calculations. In addition, in order to ensure the convergency of the atomic parameters concerned for a given MR sets, the orbital in the active set were added layer by layer as mentioned earlier. For example, five layers of correlation orbitals were added for capturing adequately the higher-order correlation under the condition of $\omega > 0.05$. This was marked as MR$_1$-$8$, in which $``8" $ is the maximum principal quantum number of the active ``s" orbitals for the corresponding MR$_1$ model. The similar regulation was used for the MR$_2$ and MR$_3$ models as well. Moreover, we should point out that in the selection of the multireference configurations we balanced the weights of the given configurations for $^3P_1^o$ and $^1P_1^o$ states since these levels were optimized on a common orbital basis set. It means that the 2s$^2$2p$^6$3s5p and 2s$^2$2p$^6$3d5p were added in the MR$_2$ set, since their weights for $^1P_1^o$ state satisfy $\omega > 0.02$. At last, the contribution of the Breit interaction was evaluated in the DF model. The calculations in this work were carried out by using the GRASP2K code~\cite{Jonsson2007,Jonsson2013}.\\

\begin{table}
\caption{The number of CSFs for the ground state 3s$^2$ in various correlation models. AO represents the active orbitals in different calculation models, and NCSF is the number of CSFs.\label{3s2AO}}
\begin{tabular}{llclccc}
\hline
\hline
Reference configurations                     & AO($n_{max}l$)&$~$& Model &$~$&NCSF&$~$\\
&&&&&$~$$J^p=0^e$&$~$\\\hline
\{2s$^2$2p$^6$3s$^2$\}                       &$~$                             & $~$ & DF              &$~$  & 1      &$~$\\
$~$                                          & \{4s, 3p, 3d, 4f\}             & $~$ & CV$_{1s}$-$4f$  & $~$ & 61     &$~$\\
$~$                                          & \{5s, 4p, 4d, 5f, 5g\}         & $~$ & CV$_{1s}$-$5g$  & $~$ & 254    &$~$\\
$~$                                          & \{6s, 5p, 5d, 6f, 6g, 6h\}     & $~$ & CV$_{1s}$-$6h$  & $~$ & 616    &$~$\\
$~$                                          & \{7s, 6p, 6d, 7f, 7g, 6h\}     & $~$ & CV$_{1s}$-$7g$  & $~$ & 1098   &$~$\\
$~$                                          & \{8s, 7p, 7d, 8f, 7g, 6h\}     & $~$ & CV$_{1s}$-$8f$  & $~$ & 1603   &$~$\\
$~$                                          & \{9s, 8p, 8d, 9f, 7g, 6h\}     & $~$ & CV$_{1s}$-$9f$  & $~$ & 2211   &$~$\\
$~$                                          & \{10s, 9p, 9d, 10f, 7g, 6h\}   & $~$ & CV$_{1s}$-$10f$ & $~$ & 2922   &$~$\\
$~$                                          & \{11s, 10p, 10d, 10f, 7g, 6h\} & $~$ & CV$_{1s}$-$11d$ & $~$ & 3509   &$~$\\
$~$                                          & \{12s, 11p, 11d, 10f, 7g, 6h\} & $~$ & CV$_{1s}$-$12d$ & $~$ & 4163   &$~$\\
$~$                                          & \{13s, 12p, 12d, 10f, 7g, 6h\} & $~$ & CV$_{1s}$-$13d$ & $~$ & 4884   &$~$\\
$~$                                          & \{13s, 12p, 12d, 10f, 7g, 6h\} & $~$ & CC$_{2s}$       & $~$ & 14988  &$~$\\
$+$\{2s$^2$2p$^6$3p$^2$\}                    & \{8s, 8p, 7d, 8f, 7g, 6h\}     & $~$ & MR$_1$-$8$      & $~$ & 23358  &$~$\\
$+$\{2s$^2$2p$^4$3s$^2$4p$^2$; 2s$^2$2p$^4$3s$^2$3d$^2$;& \{6s, 7p, 6d, 6f, 6g, 6h\}& $~$ & MR$_2$-$6$ & $~$ & 229659 &$~$\\
$~~~~$2s$^2$2p$^5$3s3p3d \}\\
$+$\{2s$^2$2p$^4$3s$^2$3p4p; 2s$^2$2p$^4$3s$^2$3d4d;    & \{5s, 6p, 5d, 4f, 5g\}& $~$ & MR$_3$-$5$    & $~$ & 357223 &$~$\\
$~~~~$2s2p$^5$3s$^2$4s4p; 2s$^2$2p$^4$3s$^2$4p5p\} &\\
\hline
\hline
\end{tabular}
\end{table}

\begin{table}
\caption{The number of CSFs for the excited states 3s3p in various correlation models. AO represents the active orbitals in different calculation models, and NCSF is the number of CSFs.\label{3s3pAO}}
\begin{tabular}{llclc}
\hline
\hline
Reference configurations& AO($n_{max}l$)&$~$& Model &NCSF\\
&&$~$&&$J^p=1^o$\\\hline
\{2s$^2$2p$^6$3s3p\}                                      & $~$                           & $~$ & DF              &2       \\
$~$                                                       & \{4s, 4p, 3d, 4f\}            & $~$ & CV$_{1s}$-$4f$  & 600    \\
$~$                                                       & \{5s, 5p, 4d, 5f, 5g\}        & $~$ & CV$_{1s}$-$5g$  & 2326   \\
$~$                                                       & \{6s, 6p, 5d, 6f, 6g, 6h\}    & $~$ & CV$_{1s}$-$6h$  & 5573   \\
$~$                                                       & \{7s, 7p, 6d, 7f, 7g, 6h\}    & $~$ & CV$_{1s}$-$7g$  & 9860   \\
$~$                                                       & \{8s, 8p, 7d, 8f, 7g, 6h\}    & $~$ & CV$_{1s}$-$8f$  & 14292  \\
$~$                                                       & \{9s, 9p, 8d, 9f, 7g, 6h\}    & $~$ & CV$_{1s}$-$9f$  & 19594  \\
$~$                                                       & \{10s, 10p, 9d, 10f, 7g, 6h\} & $~$ & CV$_{1s}$-$10f$ & 25766  \\
$~$                                                       & \{11s, 11p, 10d, 10f, 7g, 6h\}& $~$ & CV$_{1s}$-$11d$ & 30696  \\
$~$                                                       & \{12s, 12p, 11d, 10f, 7g, 6h\}& $~$ & CV$_{1s}$-$12d$ & 36146  \\
$~$                                                       & \{13s, 13p, 12d, 10f, 7g, 6h\}& $~$ & CV$_{1s}$-$13d$ & 42116  \\
$~$                                                       & \{13s, 13p, 12d, 10f, 7g, 6h\}& $~$ & CC$_{2s}$       & 223468 \\
$+$\{2s$^2$2p$^6$3p3d\}                                   & \{8s, 8p, 8d, 8f, 7g, 6h\}    & $~$ & MR$_1$-$8$      & 352702 \\
$+$\{2s$^2$2p$^6$3s5p;  2s$^2$2p$^6$3s3p; 2s$^2$2p$^6$3d5p\} & \{5s, 7p, 5d, 5f, 5g\}     & $~$ & MR$_2$-$5$      & 400690 \\
$+$\{2s$^2$2p$^4$3s3p4p$^2$; 2s$^2$2p$^4$3s3p4d$^2$\}        &\{4s, 6p, 5d, 4f \}         & $~$ & MR$_3$-$4$      & 1327012\\
\hline
\hline
\end{tabular}
\end{table}
\begin{table}
\caption{The weight factors $\omega$ of the configurations in multireference configurations (MR) for the 3s$^2$ $^1S_0$ and 3s3p $^{3,1}P^o_1$ states.\label{MR} }
\begin{tabular}{llcclcclc}
\hline
\hline
Model &\multicolumn{2}{c}{3s$^2$ $^1S_0$}&&\multicolumn{2}{c}{3s3p $^{3}P^o_1$}&&\multicolumn{2}{c}{3s3p $^{1}P^o_1$}\\\cline{2-3}\cline{5-6}\cline{8-9}
      & Configurations            & $\omega$ && Configurations          & $\omega$ && Configurations          & $\omega$  \\\hline
MR$_1$& 2s$^2$2p$^6$3s$^2$        & 0.9810   && 2s$^2$2p$^6$3s3p        & 0.9896   && 2s$^2$2p$^6$3s3p        & 0.9781    \\
      & 2s$^2$2p$^6$3p$^2$        & 0.1382   && 2s$^2$2p$^6$3p3d        & 0.0557   && 2s$^2$2p$^6$3p3d        & 0.1439    \\\hline
MR$_2$& 2s$^2$2p$^4$3s$^2$3d$^2$  & 0.0385   && 2s$^2$2p$^6$3s5p        & 0.0175   && 2s$^2$2p$^6$3s5p        & 0.0360    \\
      & 2s$^2$2p$^4$3s$^2$4p$^2$  & 0.0300   && 2s$^2$2p$^6$3s4p        & 0.0209   && 2s$^2$2p$^6$3s4p        & 0.0270    \\
      & 2s$^2$2p$^5$3s3p3d        & 0.0204   && 2s$^2$2p$^6$3d5p        & 0.0085   && 2s$^2$2p$^6$3d5p        & 0.0262    \\\hline
MR$_3$& 2s$^2$2p$^4$3s$^2$3p4p    & 0.0179   && 2s$^2$2p$^4$3s3p4d$^2$  & 0.0194   && 2s$^2$2p$^4$3s3p4d$^2$  & 0.0188    \\
      & 2s$^2$2p$^4$3s$^2$3d4d    & 0.0162   && 2s$^2$2p$^4$3s3p4p$^2$  & 0.0160   && 2s$^2$2p$^4$3s3p4p$^2$  & 0.0111    \\
      & 2s2p$^5$3s$^2$4s4p        & 0.0162   && \\
      & 2s$^2$2p$^4$3s$^2$4p5p    & 0.0158   && \\
 \hline
\hline
\end{tabular}
\end{table}

\section{Results and discussion}
\subsection{Hyperfine interaction constants of 3s3p  $^{3,1}P^o_1$}
$~$ Figure \ref{figure 1} shows the magnetic dipole ($A$) and the electric quadrupole ($B$) hyperfine interaction constants of the 3s3p $^{3,1}P^o_1$ states in Al$^+$ ions as functions of the configuration spaces. It can be seen that seven layers of correlation orbitals are enough to capture the VV and CV electron correlation effects in the SCF calculations. The extra three layers were augmented for adequately describing the core-core correlation in the subsequent RCI computation. In addition, the relatively large oscillation for the electric quadrupole hyperfine interaction constants of the 3s3p $^{1}P^o_1$ state occurs because the scale in this figure is small. The contributions of CC$_{2s}$ correlations to the $A$ and $B$ constants are $-$4\% and $-$9\% for the $^{3}P^o_1$ state, respectively, and 9\% and 5\% for the $^{1}P^o_1$ state. The higher-order correlation on the hyperfine interaction constants were accounted for by the MR$_3-4$ model. These contribute to the magnetic dipole and the electric quadrupole hyperfine interaction constants of $^{3}P^o_1$ state by 0.6\% and 3\%, respectively, and of the $^{1}P^o_1$ state by $-$8\% and $-$5\%. It is worth noting that the higher-order correlation effects counteract the core-core correlation effects. For instance, the effects of the CC$_{2s}$ correlation on the magnetic dipole hyperfine interaction constants of the $^{1}P^o_1$ states are offset by the higher-order correlation, so the final results obtained with the MR$_3-4$ model are in good agreement to the one from the CV$_{1s} - 13d$ models. In fact, Engles and Jacek {\it {et al.}} found this phenomenon in their works \cite{Bieron2009,Engels1993}, that is, the CC correlations always make the agreement worse between the calculated hyperfine interaction constants and the experimental values, and these discrepancies can be offset by the higher-order correlation. Hence, for the calculation asked for high precision, the CC correlation and the higher-order correlation are indispensable, which is also allowed us to evaluate the uncertainties in the calculation.
\begin{figure}[htb!]
\centering
\begin{minipage}{\textwidth}
\includegraphics[scale=0.25]{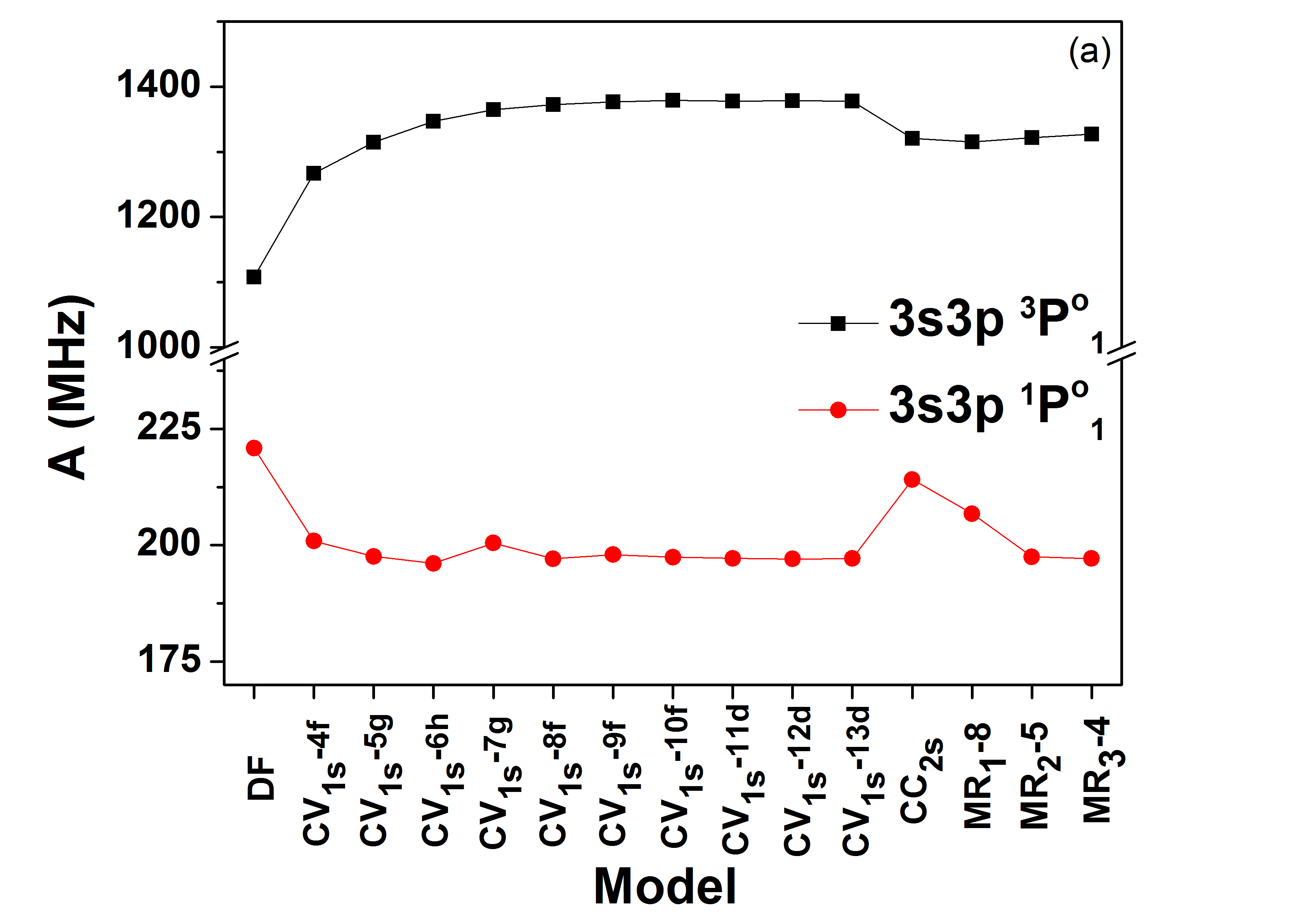}
 \includegraphics[scale=0.25]{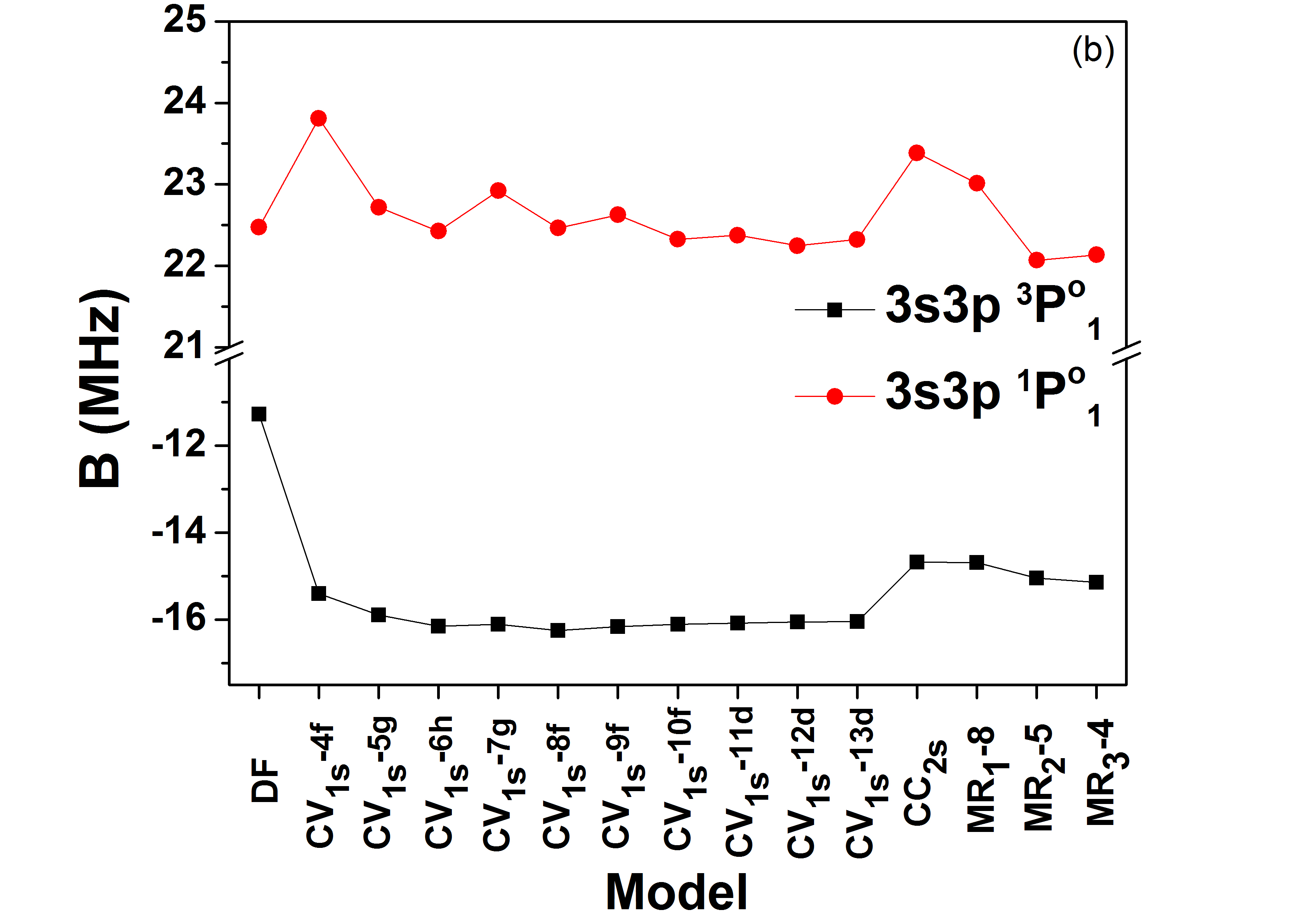}
 \caption{The hyperfine interaction constants (in MHz) $A$ (a) and $B$ (b) of 3s3p $^{3,1}P_1^o$ states in the Al$^+$ ion as functions of the computational models. \label{figure 1}}
\end{minipage}
\end{figure}

$~$ Kang {\it {et al.}} and Andersson {\it {et al.}}\cite{Kang2009,Andersson2010} calculated the off-diagonal hyperfine interaction constants using the MCDHF method in order to investigate the influence of hyperfine interaction on the lifetime of metastable states 3s3p $^3P^o_{0,2}$. For confirming our computation models are reliable, we made comparisons for the off-diagonal hyperfine interaction constants between our and their results in Table \ref{table compare}. As can be seen from this table, excellent agreement was found between ours and Andersson {\it {et al.}}. The less good agreement for Kang {\it {et al.}} results is caused by the fact that the core-valence correlation between the 1s orbitals and the valence orbitals, the core-core correlation and the higher-order correlation were not considered.

$~$ Table \ref{table AB} shows the calculated hyperfine interaction constants corresponding to the MR$_3$-$4$ model and their uncertainties (in the parenthesis) of the 3s3p $^{3,1}P^o_1$ states in Al$^+$ ions. Generally, the uncertainties result from electron correlations neglected in the computational models and physical effects. In this work the VV, CV$_{1s}$, CC$_{2s}$ and the main part of higher-order correlations were taken into account systematically. According to the convergence trend illustrated in Figure \ref{figure 2}, there are 0.8\% uncertainties in the calculations. The CC$_{1s}$ correlation and the higher-order correlation involving the 1s orbital were neglected in our computational models. From our test we found that the effects of the CC$_{1s}$ correlations on the hyperfine interaction constants under investigation are less than $1\%$. As discussed above, the effects of the core-core correlation related to the $n=2$ shell on the hyperfine interaction constants almost counteract the higher-order correlation effects. We speculated that the cancellation also arises between the CC$_{1s}$ and the related higher-order correlation. Therefore, the uncertainty due to the neglected CC$_{1s}$ and the higher-order effects should be less than 1\%. Additionally, the Breit interaction was ignored in the present calculation, which gives rise to 0.05\% uncertainty. The total uncertainty of the $A$ and $B$ is 0.76\% and 1.32\% for the $^{3}P^o_1$ state,  and 1.42\% and 1.36\% for the $^{1}P^o_1$ state.\\

\begin{table}
\caption{The off-diagonal hyperfine interaction constants (in MHz) between 3s3p $^{3,1}P^o_1$ and $^3P^o_0$ ($A_{30}$, $A_{10}$), and between 3s3p $^{3,1}P^o_1$ and $^3P^o_2$ ($A _{32}$ $B_{32}$, $A_{12}$ $B_{12}$). The results of Kang {\it {et al.}} and Andersson {\it {et al.}} in this table are converted from the hyperfine matrix elements $W_{30}$ and $W_{10}$ in the work of Kang {\it {et al.}}~\cite{Kang2009} and the reduced hyperfine interaction constants of the $3s3p$ in the work of Andersson {\it {et al.}}~\cite{Andersson2010}, respectively. The nuclear spin, the magnetic dipole and the electric quadrupole moments of $^{27}$Al, $I=5/2$, $\mu_I=3.6415069$ $\mu_N$, $Q=0.1466$ b were taken from the table by Stone\cite{Stone2005},. \label{table compare}}
\begin{tabular}{ccccccc}
\hline
\hline
Author                         & $A_{30}$ & $A_{10}$    & $A_{32}$    & $A_{12}$    & $B_{32}$  & $B_{12}$     \\\hline
This work                      & 1309     & 1027        & $-$537      & 826         & 7.50      & $-$0.0121    \\
Andersson \cite{Andersson2010} & 1349     & 1071        & $-$555      & 861         & 8.05      & $-$0.0098    \\
Kang   \cite{Kang2009}         & 1162     & 928       &&&&\\
\hline
\hline
\end{tabular}
\end{table}

 \begin{table}
 \caption{The hyperfine interaction constants (in MHz) A and B of 3s3p $^{3,1}P^o_1$ states in Al$^+$ ions. The results in the parenthesis are the uncertainties for our results.\label{table AB}}
\begin{tabular}{lccccc}
\hline
 \hline
Model &\multicolumn{2}{c}{3s3p $^3P^o_1$}&&\multicolumn{2}{c}{3s3p $^1P^o_1$}\\
        \cline{2-3}\cline{5-6}
           & A           & B           && A         & B        \\\hline

MR$_3$-$4$ & 1327.3(10.2)& $-$15.1(0.2)&& 197.1(2.8)& 22.1(0.3)\\
 \hline
 \hline
 \end{tabular}
 \end{table}

\subsection{Isotope shift factors of transitions 3s$^2$ $^1S_0$ $-$ 3s3p $^{3,1}P^o_1$}
$~$ Table \ref{table IS} shows the isotope shift factors, including the relativistic normal $\Delta K_{\rm RNMS}$ and specific mass shifts factors $\Delta K_{\rm RSMS}$, and the field shift factors $\Delta F$, for the transitions 3s$^2$ $^1S_0$ $-$ 3s3p $^{3,1}P^o_1$ as functions of computational models. From this table we can see that the mass shift factors are more sensitive to the electron correlations than the field shift factors, especially for the specific mass shift factors. For instance, from the model DF to CV$_{1s}$-$4f$, the change of the $\Delta K_{\rm RNMS}$ reaches 28\% for the transition 3s$^2$ $^1S_0$ $-$ 3s3p $^{3}P^o_1$, and 48\% for 3s$^2$ $^1S_0$ $-$ 3s3p $^{1}P^o_1$. While the $\Delta K_{\rm RSMS}$ reduced by a factor of 2 for the transition 3s$^2$ $^1S_0$ $-$ 3s3p $^{3}P^o_1$, and a factor of 6 for the transition 3s$^2$ $^1S_0$ $-$ 3s3p $^{1}P^o_1$, respectively. In contrast, the change for $\Delta F$ is marginal, less than 2\%. The high sensitivity of the SMS factors to the electron correlations is shown again when the CC$_{2s}$ correlations were included in RCI computations. The influences of the CC$_{2s}$ correlation on the $\Delta K_{\rm RNMS}$ is about 16\%, and the change is about three times for the $\Delta K_{\rm RSMS}$ of the transition 3s$^2$ $^1S_0$ $-$ 3s3p $^{1}P^o_1$. For the transition 3s$^2$ $^1S_0$ $-$ 3s3p $^{3}P^o_1$ the effects of CC$_{2s}$ correlation and related higher-order correlation on the isotope shift factors are opposite, which is similar to the case of the hyperfine interaction constants. However, the CC$_{2s}$ and related higher-order correlations both decrease the $\Delta K_{\rm RSMS}$ of the transition 3s$^2$ $^1S_0$ $-$ 3s3p $^{1}P^o_1$.

$~$ We display the relativistic and nonrelativistic NMS $\Delta K_{\rm RNMS}$ and $\Delta K_{\rm NMS}$, the SMS $\Delta K_{\rm RSMS}$ and$\Delta K_{\rm SMS}$, the FS $\Delta F$ factors and their uncertainties (in parenthesis) in Table \ref{table ISMR} for the transitions 3s$^2$ $^1S_0$ $-$ 3s3p $^{3,1}P^o_1$. In the non-relativistic approximation, the NMS factors for a given transition are proportional to its transition frequency $\nu$, {\it {i.e.}} the scaling law $\Delta K_{\rm NMS}= -\nu/1823$~\cite{Godefroid2001}. Using the experimental transition energies we obtained the non-relativistic normal mass shift factor $\Delta K_{\rm NMS} = -$615.67 (GHz u) for the 3s$^2$ $^1S_0$ $-$ 3s3p $^{3}P^o_1$ transition and $\Delta K_{\rm NMS} =-983.86$ (GHz u) for the 3s$^2$ $^1S_0$ $-$ 3s3p $^{1}P^o_1$ transition, respectively. Compared with our \textit{ab initio} calculation, the discrepancies are less than 3\%, which due not only to the neglected CC$_{1s}$ and related higher-order correlations but also to the relativistic effect in the atomic state wave functions. The linear correlation between the convergency of the calculated transition energy and the MS factors for a given transition has been found in the B$^+$, C$^-$ and Cl$^-$ ions~\cite{ljp2017,Carette2013,Carette2011}, which also occurs in the present case. We deduced from the linear correlation that the SMS factors are changed in a right direction with the MR calculations, since the transition energies in MR$_3$ model agree better with the experimental value than those in MR$_1$ and MR$_2$ models. In addition, from Table \ref{table ISMR}, it was found that for the transition 3s$^2$ $^1S_0$ $-$ 3s3p $^{3}P^o_1$ the contribution of the relativistic nuclear recoil corrections to the NMS and SMS factors are 24(GHz u) and 13(GHz u), respectively. For the other transition it has similar value. For the total MS factors of the transitions 3s$^2$ $^1S_0$ $-$ 3s3p $^{3,1}P^o_1$, the effects of the relativistic nuclear recoil corrections are not more than $4\%$, which is in our expectation that the relativistic nuclear recoil corrections for ions with $Z=13$ are small. In addition, we noticed that the effects of the relativistic nuclear recoil corrections on the NMS and SMS factors are insensitive to the electron correlations, which is illustrated in Figure \ref{figure 2}.

$~$ For the isotope shift factors, the uncertainties from the VV, CV$_{1s}$ and CC$_{2s}$ correlations reach a satisfactory level. Specifically, for the NMS factors of two transitions it is less than 1\%, and for the FS factors less than 0.1\%. The uncertainties of the SMS factors resulting from the VV, CV$_{1s}$ and CC$_{2s}$ correlations were controlled less than 2\% for the transition 3s$^2$ $^1S_0$ $-$ 3s3p $^{3}P^o_1$ and 7\% for 3s$^2$ $^1S_0$ $-$ 3s3p $^{1}P^o_1$. Since the limited computing resource, we cannot further expand the configuration space. Therefore, it is difficult to estimate accurately uncertainties of the mass shift factors in the MR computations due to neglected higher-order correlations. Roughly, the uncertainties reach around 9\% and 7\% for the NMS and SMS factors, respectively, in the 3s$^2$ $^1S_0$ $-$ 3s3p $^{3}P_1^o$ transition. For the 3s$^2$ $^1S_0$ $-$ 3s3p $^{1}P_1^o$ transition, the errors are about 2\%  for the NMS and 16\% for the SMS. Compared with the mass shift factors, the FS factors are stable with expansion of the configuration space. We estimated the uncertainties for the FS factors in the MR calculations to be around 1\% for these two transitions. Additionally, we have tested the effects of the CC$_{1s}$ correlation on the IS factors, and found that for the NMS factors the uncertainties are not more than $2\%$, but for the SMS factors they are about $7\%$ and $17\%$ for the 3s$^2$ $^1S_0$ $-$ 3s3p $^{3}P_1^o$ and 3s$^2$ $^1S_0$ $-$ 3s3p $^{1}P_1^o$ transitions, respectively. Nevertheless, the effect of the CC$_{1s}$ correlation on the FS factors is fractional (less than 1\%). The Breit interaction corrections to the isotope shift factors, estimated in the DF calculation, are less than 3\% for all physical quantities under investigation. To sum up, the total uncertainties are about 15\% for the NMS factors, 17\% for the SMS factors, and 3\% for the FS factor in 3s$^2$ $^1S_0$ $-$ 3s3p $^{1}P_1^o$ transition, and 4\%, 47\%  and 2\% for the other transition. The calculated NMS, SMS and FS factors together with their uncertainties are listed in Table \ref{table ISMR}.\\

 \begin{table}
 \caption{The $\Delta K_{\rm RNMS}$(GHz u), $\Delta K_{\rm RSMS}$(GHz u) and $\Delta F$(GHz$/$fm$^2$) factors for the transition 3s$^2$ $^1S_0$ $-$ 3s3p $^{3,1}P_1^o$ \label{table IS}}
 \begin{tabular}{lccccccc}
 \hline
 \hline
&\multicolumn{3}{c}{$^3P_1^o$ $-$ $^1S_0$}&&\multicolumn{3}{c}{$^1P^o_1$ $-$ $^1S_0$}\\
 \cline{2-4}\cline{6-8}
Model&$\Delta K_{\rm RNMS}$&$\Delta K_{\rm RSMS}$&$\Delta F$&&$\Delta K_{\rm RNMS}$&$\Delta K_{\rm RSMS}$&$\Delta F$\\\cline{2-4}\cline{6-8}
DF             &  $-$614 &  $-$844  & $-$0.1944 && $-$615  &  $-$129 & $-$0.1945   \\
CV$_{1s}$-$4f$ &  $-$788 &  $-$1646 & $-$0.1915 && $-$911  &  $-$732 & $-$0.1921   \\
CV$_{1s}$-$5g$ &  $-$621 &  $-$967  & $-$0.1965 && $-$887  &    56   & $-$0.1873   \\
CV$_{1s}$-$6h$ &  $-$647 &  $-$943  & $-$0.2025 && $-$957  &    86   & $-$0.1928   \\
CV$_{1s}$-$7g$ &  $-$622 &  $-$972  & $-$0.2009 && $-$937  &    46   & $-$0.1911   \\
CV$_{1s}$-$8f$ &  $-$633 &  $-$975  & $-$0.2029 && $-$975  &    49   & $-$0.1928   \\
CV$_{1s}$-$9f$ &  $-$632 &  $-$975  & $-$0.2028 && $-$978  &    48   & $-$0.1930   \\
CV$_{1s}$-$10f$&  $-$623 &  $-$974  & $-$0.2026 && $-$971  &    49   & $-$0.1927   \\
CV$_{1s}$-$11d$&  $-$624 &  $-$974  & $-$0.2026 && $-$977  &    50   & $-$0.1927   \\
CV$_{1s}$-$12d$&  $-$623 &  $-$974  & $-$0.2026 && $-$977  &    51   & $-$0.1927   \\
CV$_{1s}$-$13d$&  $-$624 &  $-$974  & $-$0.2026 && $-$977  &    51   & $-$0.1927   \\
CC$_{2s}$      &  $-$597 &  $-$596  & $-$0.2021 && $-$823  &  $-$152 & $-$0.1930   \\
MR$_1$-$8$     &  $-$617 &  $-$771  & $-$0.1928 && $-$841  &  $-$112 & $-$0.1890   \\
MR$_2$-$5$     &  $-$718 &  $-$1154 & $-$0.1982 && $-$1064 &  $-$452 & $-$0.1916   \\
MR$_3$-$4$     &  $-$608 &  $-$989  & $-$0.1971 && $-$931  &  $-$268 & $-$0.1904   \\
 \hline
 \hline
 \end{tabular}
 \end{table}

 \begin{table}
 \caption{The $\Delta K_{\rm NMS}$(GHz u), $\Delta K_{\rm RNMS}$(GHz u), $\Delta K_{\rm SMS}$(GHz u), $\Delta K_{\rm SMS}$(GHz u) and $\Delta F$(GHz$/$fm$^2$) factors for the transitions 3s$^2$ $^1S_0$ $-$ 3s3p $^{3,1}P^o_1$ \label{table ISMR}}
  \begin{tabular}{lccccc}
 \hline
 \hline
Transition          & $\Delta K_{\rm NMS}$&$\Delta K_{\rm RNMS}$&$\Delta K_{\rm SMS}$&$\Delta K_{\rm RSMS}$&$\Delta F$\\\hline
$^1S_0$$-$$^3P^o_1$ & $-$632(88)          & $-$608(90)          & $-$1003(163)        & $-$989(171)          & $-$0.1971(0.0059)\\
$^1S_0$$-$$^1P^o_1$ & $-$958(38)          & $-$931(37)          & $-$279(139)         & $-$268(134)          & $-$0.1904(0.0038)\\
 \hline
 \hline
 \end{tabular}
 \end{table}

\begin{figure}[htb!]
\centering
\begin{minipage}{\textwidth}
\includegraphics[scale=0.25]{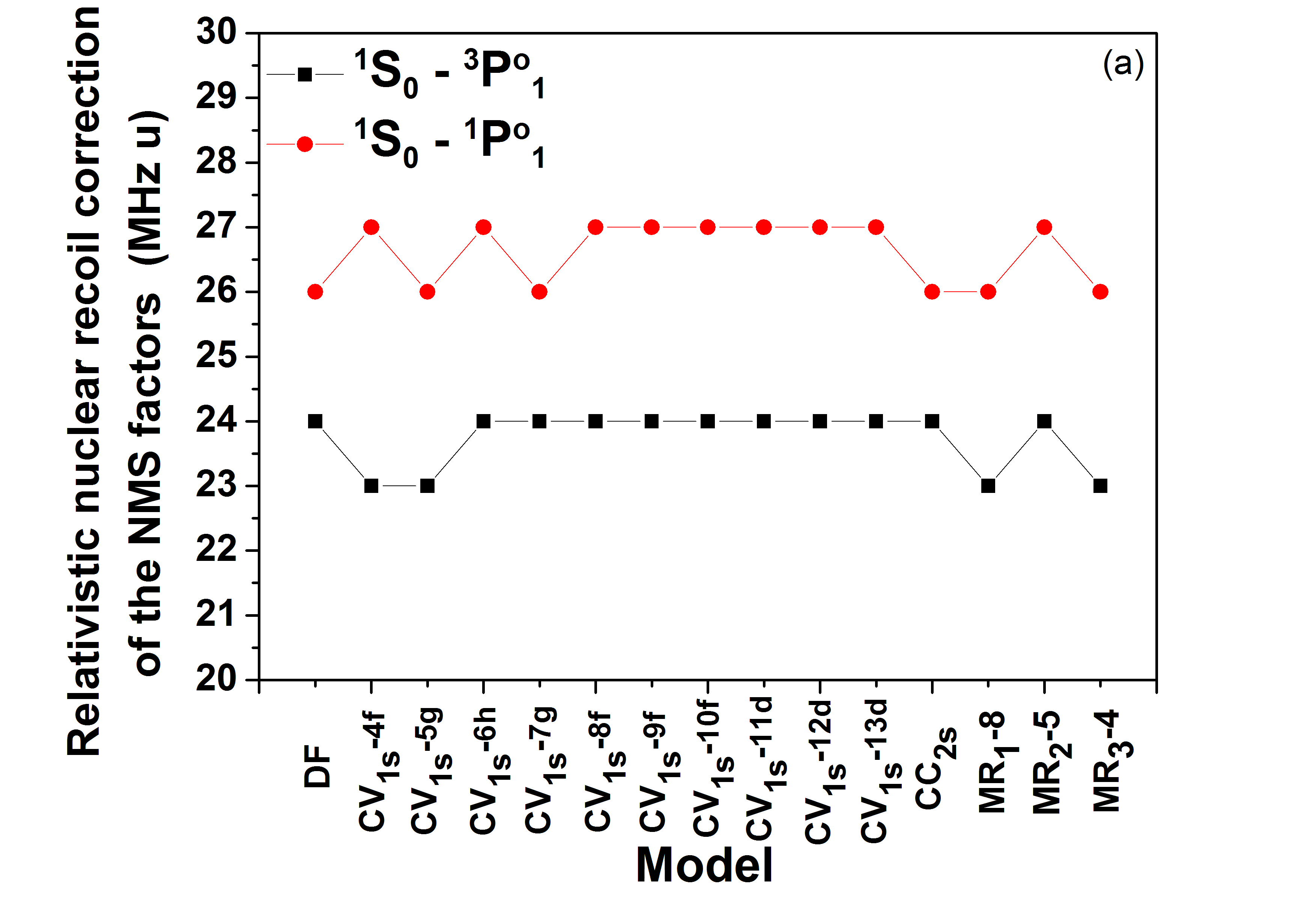}
 \includegraphics[scale=0.25]{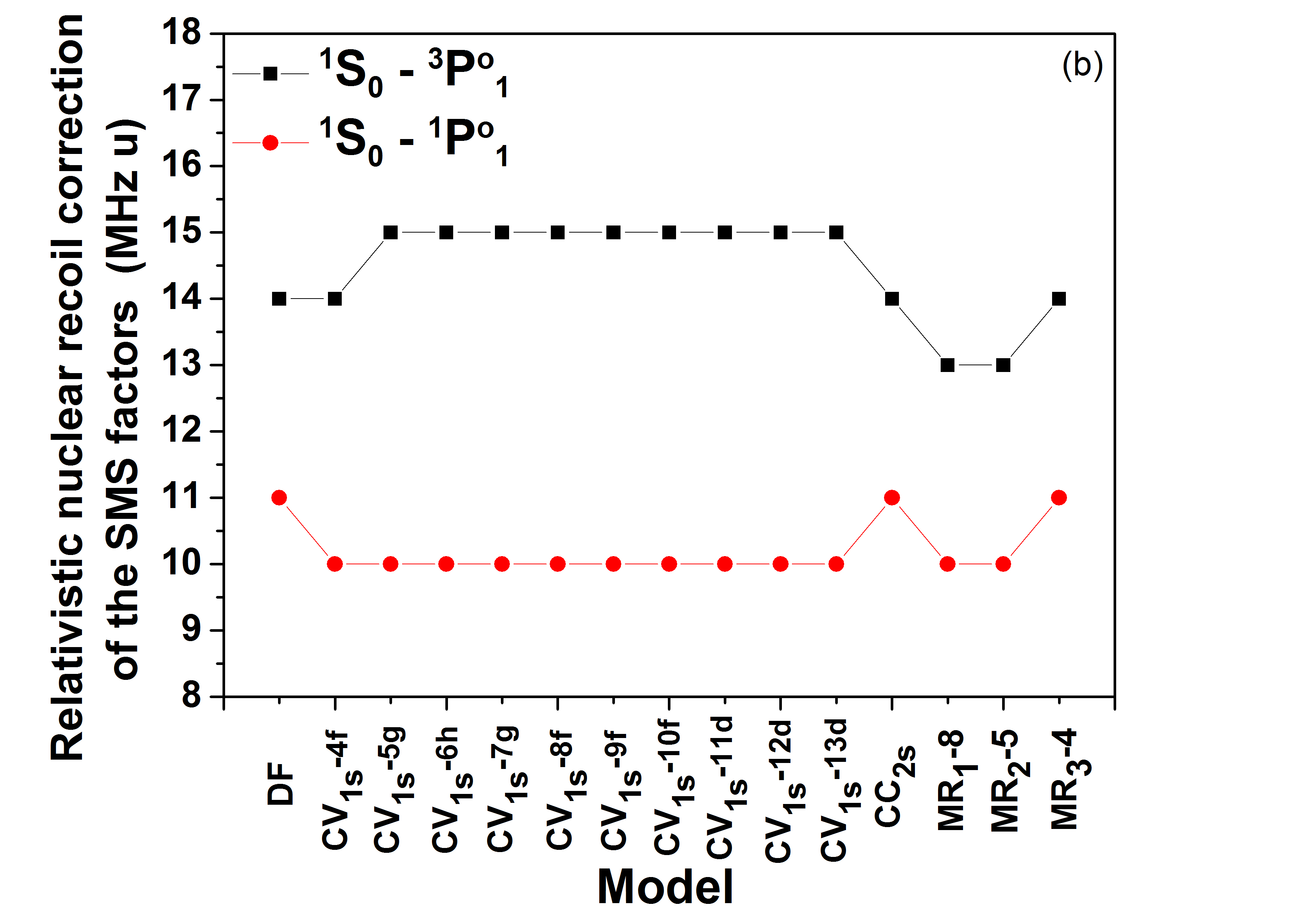}
 \caption{The relativistic nuclear recoil corrections of  the NMS (a) and SMS (b) factors for the transitions 3s$^2$ $^1S_0$ - 3s3p $^{3,1}P_1^o$ in the Al$^+$ ion as functions of the computational models. \label{figure 2}}
\end{minipage}
\end{figure}

\section{Conclusion}
In summery, we calculated the hyperfine interaction constants and the isotope shift factors involving the 3s$^2$ $^1S_0$ and 3s3p $^{3,1}P_1^o$ states in Al$^+$ ions using the MCDHF method and the active space approach. We have discussed the effects of the electron correlations and the Breit interaction on the atomic parameters concerned. In this case, we found that for the hyperfine interaction constants the contribution of the higher-order correlation is opposite with CC correlation. Based on the discussion about the contribution of the electron correlations and the Breit interaction, we have obtained reliable uncertainties for our calculated results. For the hyperfine interaction constants, the uncertainties are less than 1.5\%. For the isotope shift factors, the uncertainties are 14\% and 4\% for the NMS factors, 17\% and 47\% for the SMS factors, and 3\% and 2\% for the FS factors in transition 3s$^2~^1S_0$ $-$ 3s3p $^{3}P^o_1$ and 3s$^2~^1S_0$ $-$ 3s3p $^{1}P^o_1$, respectively. Although the higher-order correlations change the level structure slightly, these effects on the hyperfine interaction constants and the isotope shift factors are indispensable. Therefore, it is necessary to include the CC correlation and related higher-order correlation in computational models in order to achieve high accuracy of atom parameters. In addition, for the Al$^+$ ion the effect of the relativistic nuclear recoil correction on the mass shift factor is small (less than 4\%) and insensitive to the electron correlations.

With respect to the fact that there are relatively large uncertainties in the calculation of mass shift factors, the partitioned correlation function interaction (PCFI) approach~\cite{verdebout2013} will be a promising method for calculating the isotope shifts more accurately, which can capture the different electron correlations flexible and accurate.

During the review of our article, a paper about hyperfine-mediated electric quadrupole shifts in Al$^+$ and In$^+$ ion clocks was just published~\cite{Beloy2017}. In this paper, the diagonal magnetic dipole and electric quadrupole hyperfine interaction matrix elements of the 3s3p $^{3}P^o_1$ state for Al$^+$ were calculated by using the method of configuration interaction plus many-body perturbation theory (CI+MBPT). For the diagonal magnetic dipole hyperfine interaction matrix element, our value is in perfect agreement with theirs. While the consistence is less good for the diagonal electric quadrupole hyperfine interaction matrix element, and the difference is around 35\%.

%
%
%
\begin{acknowledgments}
 This work has been supported by the Nation Natural Science Foundation of China under Grant Nos.11404025, 91536106, U1331122 and U1530142, the China Postdoctoral Science Foundation under Grant No.2014M560061, the Young Teachers Scientific Research Ability Promotion Plan of Northwest Normal University (NWNU-LKQN-15-3).
\end{acknowledgments}

\end{document}